\documentclass[aps,prl,twocolumn, article]{revtex4}
\pdfoutput=1

\usepackage{graphicx}
\usepackage{amsmath, amssymb}
\usepackage{siunitx}
\usepackage{upgreek}

\newcommand{\figu}[1]{Figure~\ref{Fig:#1}}

\newcommand{\dc}{$~\upmu$m$^2$~s$^{-1} $}
\newcommand{\x}{\textit{x}}
\newcommand{\y}{\textit{y}}
\newcommand{\z}{\textit{z}}

\begin{document}

\title{Isotropic diffusion of the small ribosomal subunit in \textit{Escherichia coli}}

\author{Arash Sanamrad\footnote{A. Sanamrad and F. Persson contributed equally to this work.}}
\author{Fredrik Persson$^*$}
\author{Johan Elf\footnote{Corresponding Author: johan.elf@icm.uu.se}}
\affiliation{Department of Cell and Molecular Biology, Science for Life Laboratory, Uppsala University, Sweden}

\date{\today}

\begin{abstract}
The ribosome is one of the most important macromolecular complexes in a living cell. We have tracked individual 30S ribosomal subunits in exponentially growing \textit{E. coli} cells using three-dimensional single-particle tracking, where the \textit{z}-position is estimated using astigmatism. The 30S subunits are stochiometrically labeled with S2-mEos2 by replacing the \textit{rpsB} gene in the \textit{E. coli} chromosome by \textit{rpsB-mEos2}. The spatial precision in tracking is 20~nm in \textit{xy} and 70~nm in \textit{z}. The average trajectory consists of 4 steps corresponding to 80~ms. The trajectories are excluded from parts of the cell, consistent with nucleoid exclusion, and display isotropic diffusion with nearly identical apparent diffusion coefficients (0.05~$\pm$~0.01\dc) in \x, \y~and \z. The tracking data fits well to a two-state diffusion model where 46\% of the molecules are diffusing at 0.02\dc~and 54\% are diffusing at 0.14\dc. These states could correspond to translating 70S ribosomes and free 30S subunits.
\end{abstract}

\maketitle

Fluorescence microscopy has revolutionized cell biology during the last decades. However, a fundamental limitation of using optical microscopy in subcellular studies in bacteria has been the inherent resolution limitation of approximately 200~nm, imposed by diffraction. This makes it impossible to distinguish two objects that are closer than 200~nm. For fluorescence microscopy, diffraction-unlimited resolution is achievable by inhibiting the simultaneous fluorescence of fluorophores within a diffraction-limited area, \textit{i.e.} by spacing the observations in time instead of space. One group of techniques are capable of achieving this by stochastically allowing fluorescence from a small subset of fluorophores and are commonly referred to as (F)PALM or STORM \cite{Rust2006, Hess2006}. These techniques rely on the minute probability of activating multiple fluorophores within a diffraction-limited area. To find the spatial position centroid calculation has to be performed and this is most commonly achieved by fitting a two-dimensional Gaussian function to the point spread function (PSF). Since these stochastic methods are inherently easy to integrate in a normal wide-field fluorescence microscope with basically only a strict requirement of fluorophores that can be converted or switched on and off, they are widely used in biology. Since the development of photoconvertible fluorescent proteins, biological applications for super-resolution imaging have started to appear and are becoming increasingly popular due to their non-invasive nature \cite{Huang2010}. Lately these techniques have expanded from being inherently two-dimensional to yielding position data in all three dimensions. This is most commonly achieved by either introducing an optical 'defect', astigmatism, in form of a cylindrical lens in front of the camera (astigmatism approach) \cite{Huang2008}, or by splitting the fluorescence and detecting it on two different parts of the camera with one having a longer light path, leading to images of different focal planes (biplane approach) \cite{Juette2008}. More exotic versions based on for example interferometry \cite{Shtengel2009}, 4Pi setups \cite{Nagorni1998} and complex PSF shaping setups relying on \textit{e.g.} adaptive optics \cite{Pavani2009} are also available.

Single-particle tracking (SPT) with nanometer precision ($\sim$10 nm) for single molecules was initially demonstrated by tracking individual lipids in a lipid membrane \cite{Schmidt1996}. The particle tracking was extended to the third dimension by Kao and Verkman who introduced three-dimensional (3D) tracking using astigmatism \cite{Kao1994}. Another step forward was taken by Manley \textit{et al}. and Niu \textit{et al.} who combined SPT with stochastic super-resolution techniques by using photoconvertible proteins \cite{Manley2008, Niu2008}. This made it possible to work with abundant molecules by converting a few molecules at a time. When the molecules bleach more molecules can be converted and tracked in the same cell until a sufficient number of trajectories have been acquired. Here we have combined tracking of photoconvertible proteins with 3D localization to study the diffusion of the small ribosomal subunit labeled with the photoconvertible protein mEos2 fused to the ribosomal protein S2. 

\section{Results and Discussion}

\begin{figure}
  \includegraphics[width=\columnwidth,keepaspectratio]{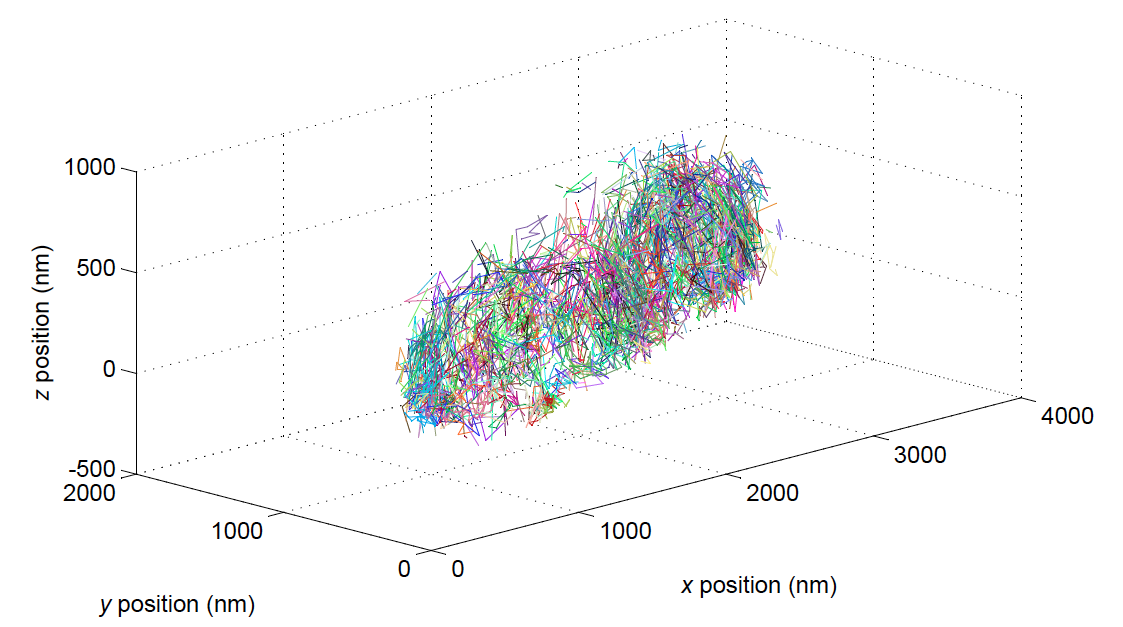}
  \centering
  \caption{Overlay of 1222 S2-mEos2 trajectories obtained from one \textit{E. coli} cell. The trajectories form a cylinder with hemispherical end caps, which is the typical shape of an \textit{E. coli} cell.} 
  \label{Fig:1}
\end{figure}

\begin{figure}
  \includegraphics[width=\columnwidth,keepaspectratio]{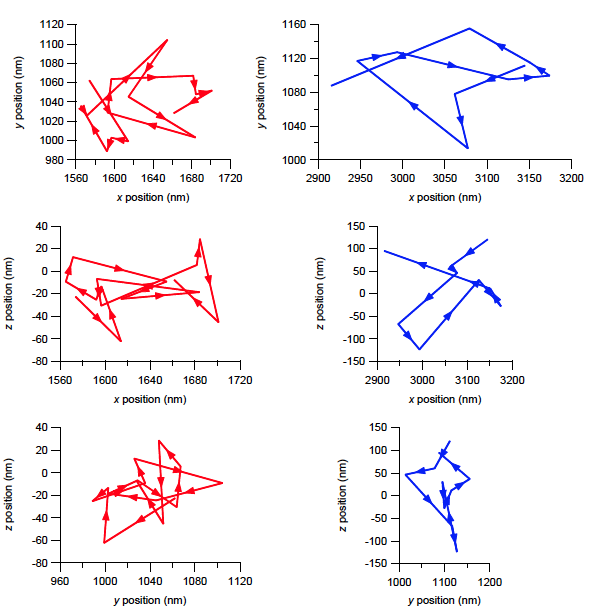}
  \centering
  \caption{Two S2-mEos2 trajectories obtained from one \textit{E. coli} cell. The trajectories are confined to regions with radii of approximately 100 and 150~nm.} 
  \label{Fig:2}
\end{figure}

We obtained 1222 S2-mEos2 trajectories from one \textit{Escherichia coli} (\textit{E. coli}) cell (see \figu{1}). The trajectories form a cylinder with hemispherical end caps, which is the typical shape of an \textit{E. coli} cell. \figu{2} shows two individual S2-mEos2 trajectories which are confined to regions with radii of approximately 100 and 150~nm. This confinement is most likely due to the tethering of the 30S subunit to mRNAs  \cite{English2011}.

\begin{figure}
  \includegraphics[width=\columnwidth,keepaspectratio]{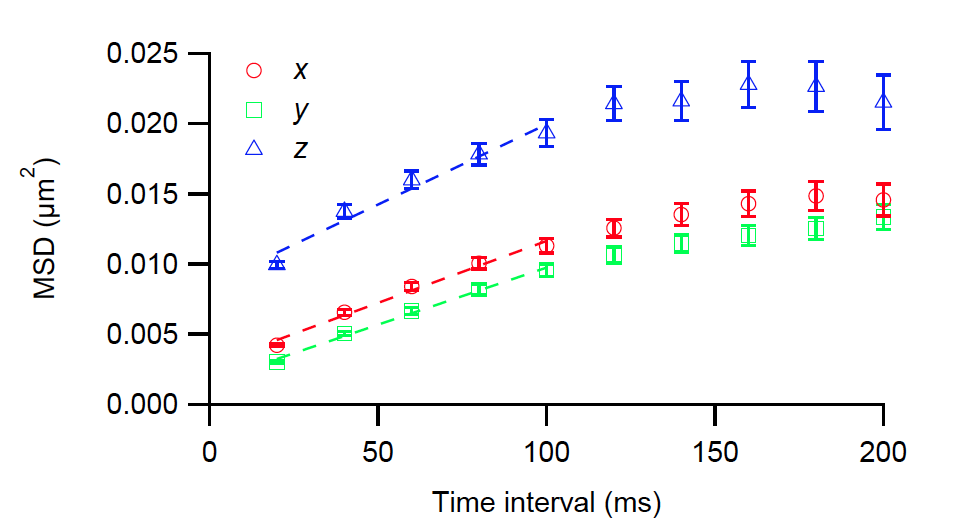}
  \centering
  \caption{Mean square displacements (MSDs) of the S2-mEos2 trajectories shown in \figu{1}. Error bars represent standard error of the means. All MSD curves plateau after 200~ms. The \x~and \y~MSDs are nearly identical while the \z~MSD curve has a higher offset than the \x~and \y~MSD curves but is otherwise very similar to these curves. The \x, \y~and \z~apparent diffusion coefficients obtained by fitting lines to the first 5 MSDs are 0.04, 0.04 and 0.06\dc, respectively.} 
  \label{Fig:3}
\end{figure}

To investigate whether the diffusion of the small subunit is isotropic or not, we calculated the \x, \y~and \z~mean square displacements (MSDs) of the trajectories (see \figu{3}). The  MSD curves plateau after 200~ms and the \x~and \y~MSDs are nearly identical, indicating that cellular confinement has a very small effect on S2-mEos2 diffusion at this timescale. The \z~MSD curve has a higher offset than the \x~and \y~MSD curves but is otherwise very similar to these curves. The higher offset can be explained by the higher uncertainties of the \textit{z}-positions. The \x, \y~and \z~apparent diffusion coefficients obtained by fitting lines to MSDs up to 100~ms are 0.04, 0.04 and 0.06\dc, respectively. Unlike the MSDs themselves, these coefficients are very similar since they are largely unaffected by the localization errors which mainly affect the offsets.  

\begin{figure}
  \includegraphics[width=\columnwidth,keepaspectratio]{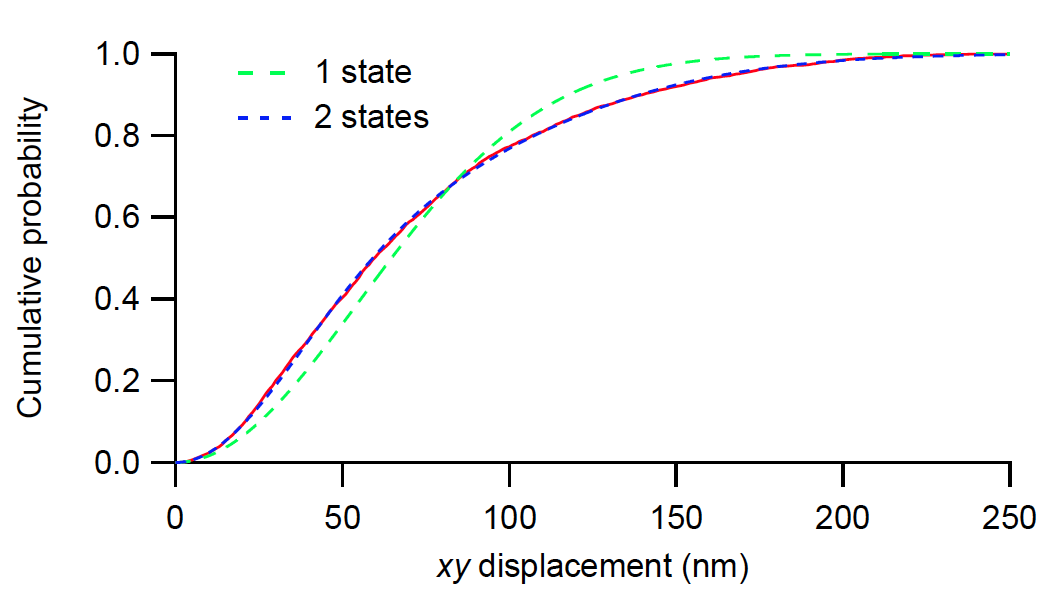}
  \centering
  \caption{Cumulative distribution function (CDF) of the \textit{xy} displacements over 20~ms, calculated from the S2-mEos2 trajectories shown in \figu{1}. The CDF is fitted with two CDF models of Brownian motion corresponding to one and two diffusion states. The sums of the squared errors are 0.4 and 0.004, respectively.} 
  \label{Fig:4}
\end{figure}

To further analyze the trajectories, we fitted cumulative distribution functions (CDFs) corresponding to one-, two-, and three-state diffusion models to the experimental CDF of 20-ms displacements (see \figu{4}). It is clear that a one-state model is not sufficient in describing the data. The sum of the squared errors decrease two orders of magnitude when the number of states is increased from one to two but only a factor of two when the number of states is increased from two to three. This indicates that a model with two diffusion states is sufficient to explain the data. The apparent diffusion coefficient obtained from fitting the experimental CDF with a CDF corresponding to a one-state diffusion model is 0.07\dc. This apparent diffusion coefficient is higher than the apparent diffusion coefficients obtained from the initial slopes of the MSD curves, $0.05~\pm~0.01$\dc, since it also includes the localization errors. 

The apparent diffusion coefficients obtained from fitting the experimental CDF with a CDF corresponding to two diffusion states are $0.14$\dc~for the fast state and $0.03$\dc~for the slow state. When we take the localization errors into account we obtain $D_{\text{1}} = 0.14$\dc~and $D_{\text{2}} = 0.02$\dc. The occupancies of the two states are 54\% for the fast state and 46\% for the slow state. The diffusion states probably correspond to free 30S ribosomal subunits and translating 70S ribosomes. If this is the case, it implies that 54\% of the small subunits are involved in translation under our experimental conditions. It should be noted that free S2-mEos2 proteins in the cell diffuse too fast to be localized with 20-ms exposures. 

\begin{figure}
  \includegraphics[width=0.8\columnwidth,keepaspectratio]{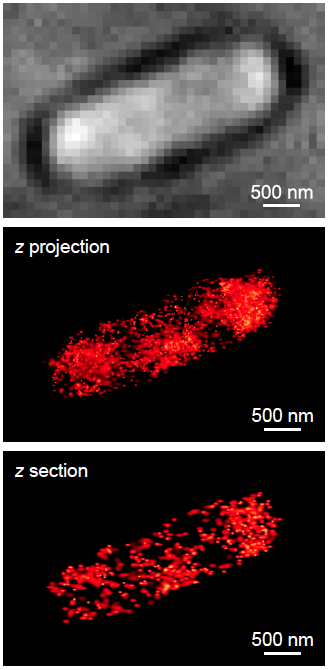}
  \centering
  \caption{Bright-field image of an \textit{E. coli} cell expressing S2-mEos2, a super-resolution image constructed from all independent S2-mEos2 localizations in the cell and a super-resolution image constructed from all independent S2-mEos2 localizations in a 200-nm \textit{z} section of the cell. Note that S2-mEos2 is excluded from certain parts of the cell, which is consistent with nucleoid exclusion.} 
  \label{Fig:5}
\end{figure}

The 3D tracking reveals that the ribosomes move isotropically in \x, \y~and \z. This is advantageous since it implies that most information about ribosome movement can be extracted from diffusion along the bacterial long axis (\x) alone. The advantage of studying diffusion along the long axis is that the diffusion in \y~and \z~will be spatially restricted by the membrane and possibly also the nucleoid. Furthermore, the confinement in \y~will depend on the \textit{z}-position, which will always be the most inaccurate in a 3D imaging system based on astigmatism.

To check if the small subunit is excluded from parts of the cell, we constructed a super-resolution image from all independent S2-mEos2 localizations in a 200-nm thick \textit{z}-section of the cell (see \figu{5}). S2-mEos2 is excluded from certain parts of the cell, consistent with nucleoid exclusion. This effect has previously been observed by electron microscopy of fixed cells \cite{Robinow1994} and super-resolution imaging of Sra-mEos2 \cite{Wang2011}.

An ideal use of single-molecule tracking data is the determination of rates by which individual molecules go between different complexes with different diffusion properties. In our case, such states could for example correspond to free small subunits and translating 70S ribosomes. However, in our case, the average trajectory consists of 4 steps, corresponding to 80~ms, which implies that we would expect that less than 1\% of the tracked ribosomes complete a round of translation within a trajectory, assuming that it would take at least 10~s to translate a typical open reading frame \cite{BremerBook}. The number of translation initiation events will be just as unlikely in a steady-state situation. Therefore, in order to study the dynamics of translation initiation, elongation, termination and ribosome recycling using single-molecule tracking, the trajectories would have to span longer times. One way of achieving this is by spacing out the imaging events such that the molecules can be monitored for longer times at the same total exposures, since photobleaching is often the limiting factor. The disadvantage of spacing out the observation time points is that it is less certain that the same molecule is being tracked, \textit{i.e.} the molecule that is being imaged can photobleach in one frame while another molecule spontaneously photoconverts before the next imaging event. The rate of spontaneous photoconversion of an mEos2 molecule is not known but can be considered to be very low. However, since there are so many ribosomes per cell ($\sim$25000) \cite{BremerBook}, the overall risk of spontaneous conversion under a time period corresponding to a translation cycle might be significant. 

To conclude, the 30S subunits are excluded from parts of the cell, consistent with nucleoid exclusion, the average apparent diffusion coefficient is 0.05\dc, and it appears that S2-mEos2 diffuses in at least two different complexes that are likely to be free 30S subunits and translating ribosomes.

\section{Methods}

\subsection{Optical setup}
The optical setup included a 405-nm photoconversion laser (Radius 405-50, Coherent), a 555-nm excitation laser (GCL-150-555, CrystaLaser), a cylindrical lens (LJ1516RM-A, Thorlabs), an EMCCD camera (DU-897E-CS0-\#BV, Andor Technology), an emission filter (HQ605/75m, Chroma Technology), an excitation filter (Z550/20x, Chroma Technology), an image splitter (Optosplit II, Cairn Research), an inverted microscope (Eclipse Ti-E, Nikon Instruments), two dichroic mirrors (Z405RDC and Z555RDC, Chroma Technology), two shutter drivers (T132 and VMM-D3, Vincent Associates) and two shutters (LS6T2 and LS6ZM2, Vincent Associates). MetaMorph 7.7.6.0 (Molecular Devices) was used to control the LS6T2 shutter, the camera and the microscope.

\subsection{Calibration of photon numbers}
The baseline offset and the standard deviation of the camera noise were determined by acquiring 200 images with 256~x~256 pixels at 50~Hz with a closed shutter, and calculating and fitting a Gaussian function to the experimental probability mass function of the intensities. The multiplication factor was determined by acquiring 200 images with 256~x~256 evenly illuminated pixels at 50~Hz with a white paper placed in front of the objective, flattening them to correct for any unevenness in the illumination, and calculating and fitting a theoretical probability mass function of intensities to the experimental probability mass function of the intensities \cite{Ulbrich2007}. The baseline offset, the multiplication factor and the quantum efficiency of the camera were used to convert the image intensities from counts to number of photons.

\subsection{Calibration of \textit{z}-position}
Fluorescent 40-nm beads (F8794, Molecular Probes) were attached to a Poly-L-lysine-treated coverslip which was placed on a 2.5\% agarose pad (SeaPlaque GTG Agarose, Lonza) containing M9 minimal medium supplemented with 0.4\% glucose. Individual beads were imaged at \textit{z}-positions ranging from -700~nm to 700~nm with 50-nm increments. Point spread functions (PSFs) were detected with a wavelet segmentation algorithm and fitted with two-dimensional Gaussian functions. Calibration curves were obtained by fitting cubic functions to the \x~and \y~widths of six beads.

\subsection{Strain construction}
A strain expressing S2-mEos2 with a glycine linker was constructed by lambda Red recombination \cite{Datsenko2000} of a \textit{rpsB-mEos2} DNA fragment in BW25993 containing the pKD46 plasmid. The construct was transferred to a clean BW25993 background by P1 phage transduction and the chloramphenicol resistance cassette was removed by expressing FLP from pCP20.

\subsection{Sample preparation and fluorescence imaging}
The cells were grown overnight in M9 minimal medium supplemented with 0.4\% glucose and RPMI 1640 amino acids (R7131, Sigma-Aldrich) at 37$^{\circ}$C, diluted 1:1000 in the morning and grown in the same manner to an optical density at 600~nm of 0.2 and placed on a 2.5\% agarose pad containing fresh medium. The cells were incubated for 30~min at room temperature (21$^{\circ}$C) and subsequently imaged for 2~min at 50~Hz. The laser power density of the excitation laser (555-nm) was 1~kW~cm$^{-2}$.

\subsection{Image analysis}
PSFs of S2-mEos2 molecules were identified with a wavelet segmentation algorithm \cite{Izeddin2012} and fitted with two-dimensional Gaussian functions. For each detected molecule the localization uncertainty was determined by the fitting error while the number of photons was determined by calculating the volume under the fitted Gaussian surface. The \textit{z}-position was determined by minimizing the norm given in \cite{Huang2008}. Points were accepted if the fit was successful, the norm was less or equal to 5, the number of background photons was greater than zero, the number of photons within the Gaussian fit was at least 100, the standard uncertainty of the \textit{x}-position was less or equal to 60~nm, the standard uncertainty of the \textit{y}-position was less or equal to 40~nm and the \textit{z}-position was between -200 and 700~nm.

\subsection{Trajectory construction}
The trajectories were constructed such that they only included points that could not be part of multiple trajectories. The maximum allowed \x, \y~and \z~displacements were 200~nm, 200~nm and 300~nm, respectively.

\section{Acknowledgements}
We thank Prune Leroy for helpful discussions and comments on the manuscript. This work was supported by Vetenskapsr\aa det, the G\"oran Gustafsson Stiftelse, the Knut and Alice Wallenberg Foundation and the Foundation for Strategic Research. The authors declare that they have no competing financial interests.

\section{Author contributions}
J.E. designed the research. A.S. constructed the bacterial strain. A.S. and F.P. built the optical setup. A.S. performed the experiments, analyzed the images and did the MSD analysis. F.P. did the CDF analysis. A.S., F.P. and J.E. discussed the analysis methods and the results, and wrote the paper.

\bibliography{}

\begin{thebibliography}{10}

\bibitem{Rust2006}
M.~J. Rust, M.~Bates, and X.~Zhuang.
\newblock Sub-diffraction-limit imaging by stochastic optical reconstruction
  microscopy ({STORM}).
\newblock {\em Nat. Methods}, 3(10):793--796, 2006.

\bibitem{Hess2006}
S.~T. Hess, T.~P. Girirajan, and M.~D. Mason.
\newblock Ultra-high resolution imaging by fluorescence photoactivation
  localization microscopy.
\newblock {\em Biophys. J.}, 91(11):4258--4272, 2006.

\bibitem{Huang2010}
Bo~Huang, Hazen Babcock, and Xiaowei Zhuang.
\newblock Breaking the diffraction barrier: super-resolution imaging of cells.
\newblock {\em Cell}, 143(7):1047--1058, 2010.

\bibitem{Huang2008}
B.~Huang, W.~Wang, M.~Bates, and X.~Zhuang.
\newblock Three-dimensional super-resolution imaging by stochastic optical
  reconstruction microscopy.
\newblock {\em Science}, 319(5864):810--813, 2008.

\bibitem{Juette2008}
M.~F. Juette, T.~J. Gould, M.~D. Lessard, M.~J. Mlodzianoski, B.~S. Nagpure,
  B.~T. Bennett, S.~T. Hess, and J.~Bewersdorf.
\newblock Three-dimensional sub--100 nm resolution fluorescence microscopy of
  thick samples.
\newblock {\em Nat. Methods}, 5(6):527--529, 2008.

\bibitem{Shtengel2009}
G.~Shtengel, J.~A. Galbraith, C.~G. Galbraith, J.~Lippincott-Schwartz, J.~M.
  Gillette, S.~Manley, R.~Sougrat, C.~M. Waterman, P.~Kanchanawong, M.~W.
  Davidson, R.~D. Fetter, and H.~F. Hess.
\newblock Interferometric fluorescent super-resolution microscopy resolves {3D}
  cellular ultrastructure.
\newblock {\em Proc. Natl. Acad. Sci. U. S. A.}, 106(9):3125--3130, 2009.

\bibitem{Nagorni1998}
M.~Nagorni and S.~W. Hell.
\newblock {4Pi}-confocal microscopy provides three-dimensional images of the
  microtubule network with 100- to 150-nm resolution.
\newblock {\em J. struct. Biol.}, 123(3):236--247, 1998.

\bibitem{Pavani2009}
S.~R. Pavani, M.~A. Thompson, J.~S. Biteen, S.~J. Lord, N.~Liu, R.~J. Twieg,
  R.~Piestun, and W.~E. Moerner.
\newblock Three-dimensional, single-molecule fluorescence imaging beyond the
  diffraction limit by using a double-helix point spread function.
\newblock {\em Proc. Natl. Acad. Sci. U. S. A.}, 106(9):2995--2999, 2009.

\bibitem{Schmidt1996}
T.~Schmidt, G.~J. Schütz, W.~Baumgartner, H.~J. Gruber, and H.~Schindler.
\newblock Imaging of single molecule diffusion.
\newblock {\em Proc. Natl. Acad. Sci. U. S. A.}, 93(7):2926--2929, 1996.

\bibitem{Kao1994}
H.~P. Kao and A.~S. Verkman.
\newblock Tracking of single fluorescent particles in three dimensions: use of
  cylindrical optics to encode particle position.
\newblock {\em Biophys. J.}, 67(3):1291--1300, 1994.

\bibitem{Manley2008}
S.~Manley, J.~M. Gillette, G.~H. Patterson, H.~Shroff, H.~F. Hess, E.~Betzig,
  and J.~Lippincott-Schwartz.
\newblock High-density mapping of single-molecule trajectories with
  photoactivated localization microscopy.
\newblock {\em Nat. Methods}, 5(2):155, 2008.

\bibitem{Niu2008}
L.~Niu and J.~Yu.
\newblock Investigating intracellular dynamics of {FtsZ} cytoskeleton with
  photoactivation single-molecule tracking.
\newblock {\em Biophys. J.}, 95(4):2009--2016, 2008.

\bibitem{English2011}
B.~P. English, V.~Hauryliuk, A.~Sanamrad, S.~Tankov, N.~H. Dekker, and J.~Elf.
\newblock Single-molecule investigations of the stringent response machinery in
  living bacterial cells.
\newblock {\em Proc. Natl. Acad. Sci. U. S. A.}, 108(31):E365, 2011.

\bibitem{Robinow1994}
C~Robinow and E~Kellenberger.
\newblock The bacterial nucleoid revisited.
\newblock {\em Microbiol. Rev.}, 58(2):211--232, 1994.

\bibitem{Wang2011}
W.~Wang, G-W.~W. Li, C.~Chen, X.~S. Xie, and X.~Zhuang.
\newblock Chromosome organization by a nucleoid-associated protein in live
  bacteria.
\newblock {\em Science}, 333(6048):1445--1449, 2011.

\bibitem{BremerBook}
H.~Bremer and P.~Dennis.
\newblock Modulation of chemical composition and other parameters by growth
  rate.
\newblock In Fredrick~C. Neidhardt, editor, {\em Echerichia coli and
  Salmonella: Cellular and Molecular Biology}. ASM press, Washington, 1996.

\bibitem{Ulbrich2007}
Maximilian~H. Ulbrich and Ehud~Y. Isacoff.
\newblock Subunit counting in membrane-bound proteins.
\newblock {\em Nat. Methods}, 4:319--321, 2007.

\bibitem{Datsenko2000}
K.~A. Datsenko and B.~L. Wanner.
\newblock One-step inactivation of chromosomal genes in \textit{Escherichia
  coli} {K-12} using {PCR} products.
\newblock {\em Proc. Natl. Acad. Sci. U. S. A.}, 97(12):6640--6645, 2000.

\bibitem{Izeddin2012}
I.~Izeddin, J.~Boulanger, V.~Racine, C.G. Specht, A.~Kechkar, D.~Nair,
  A.~Triller, D.~Choquet, M.~Dahan, and J.B. Sibarita.
\newblock Wavelet analysis for single molecule localization microscopy.
\newblock {\em Opt. Express}, 20(3):2081--2095, 2012.

\end{thebibliography}
\end{document}